\documentclass[11pt]{article}
\usepackage{epsfig,amscd,amssymb}
\begin{document}

\title{Phase transitions in one dimension and less}

\author{Paul Fendley$^1$ and Oleg Tchernyshyov$^2$\\
\cr
$^1$ Department of Physics\\
University of Virginia\\
Charlottesville, VA 22904-4714\\
{\tt fendley@virginia.edu}
\\
\\
$^2$  Department of Physics\\
Princeton University\\
Princeton, NJ 08544\\
{\tt otcherny@princeton.edu}
}
\maketitle

\begin{abstract}

Phase transitions can occur in one-dimensional classical statistical
mechanics at non-zero temperature when the number of components $N$
of the spin is infinite.
We show how to solve such magnets in one dimension for any $N$,
and how the phase transition develops at $N=\infty$. We discuss
$SU(N)$ and $Sp(N)$ magnets, where the transition is second-order.
In the new high-temperature phase, the correlation length is zero. We also
show that for the $SU(N)$ magnet on exactly three sites with
periodic boundary conditions, the transition becomes first order.

\end{abstract}

\section{Introduction}

It has long been known that phase transitions are uncommon in
one-dimensional classical statistical mechanics. An old argument by
Peierls shows that in models at non-zero temperature with local
interactions and a finite number of degrees of freedom, order is not
possible: the entropy gain from disordering the system will always
dominate the energy loss.  There are (at least) three ways of
avoiding this argument. The first two are well understood.  A system
at zero temperature can of course order: the system just sits in its
ground state. A system with long-range interactions can have an
energy large enough to dominate the entropy. In this
paper, we will discuss in depth a third way of obtaining a phase
transition in one dimension. This is to study systems with an infinite
number of degrees of freedom per site. 

In particular, we will study magnets with $O(N),\ SU(N)$ and $Sp(N)$
symmetry. We will see that there can be a phase transition in the
$N\to\infty$ limit. We solve these one-dimensional classical systems
for any $N$, and show how the transition occurs only in this limit;
for finite $N$ all quantities depend on the temperature analytically.
The infinite number of degrees of freedom has roughly the same effect
of increasing the effective dimensionality, but the phase transition
is very different from those in higher dimension. It is not a phase
transition between an ordered phase and a disordered one, but rather
between a disordered phase and a seriously-disordered one. In the
seriously-disordered phase, the system behaves as if it were at
infinite temperature.  The entropy has dominated the energy to the
point where the energy term does not affect the physics; each spin is
effectively independent. The infinite number of degrees of freedom
means that this serious disorder is possible even at finite
temperature.

The paper is a companion to one by Tchernyshyov and Sondhi
\cite{TS}. There it is shown that in some magnets, a mean-field
calculation yields a phase transition in any dimension. Since
mean-field results are exact at $N\to\infty$, this predicts the phase
transition we observe here. Their computation also predicts that there
is a first-order phase transition for the $SU(N\to\infty)$ magnet on just
three sites with periodic boundary conditions. Remarkably, this
first-order transition happens only for precisely three sites; for any
other number of sites greater than 1 there is a second-order transition.

It has long been known that phase transitions can occur as
$N\to\infty$ in zero-dimensional matrix models \cite{GW}.  Phase
transitions in one dimension at infinite $N$ were studied in
\cite{Hikami,Sokal}.  In particular, the largest eigenvalue for the
$SU(N)$ case discussed here was computed in \cite{Sokal} for any
$N$. Here will develop the necessary techniques systematically, and
extend these results in several ways. We explicitly find all the
eigenvalues of the transfer matrix for these magnets. All these
results are completely analytic in $N$ and in the inverse temperature
$\beta$ as long as $N$ is finite. The singularity and a phase
transition can develop when $N\to\infty$ and $\beta\to\infty$ with
$\beta/N$ remaining finite.  Knowing all the eigenvalues and their
multiplicities explicitly for any $N$ lets us show that there can be a
phase transition as $N\to\infty$ even for a finite number of sites in
one dimension.

In section 2, we find all the eigenvalues (and their
multiplicities) of the transfer matrices in a variety of one-dimensional magnets.  In section
3, we use these results to study the phase transitions which occur as
the number of sites and $N$ go to infinity.  Most of these phase
transitions are ferromagnetic, but one is antiferromagnetic.
In section 4, we study the first-order
transition for the three-site $SU(N)$ chain.
In an appendix we collect some useful mathematical results.

\section{Solving one-dimensional magnets at any $N$}

\subsection{The rotor}

To illustrate the procedure, we start with a simple rotor, the
classical XY model in one dimension. The spin is defined by a
periodic variable $0\le \theta < 2\pi$, and spins $\theta$ and $\theta'$
on adjacent sites have energy
\begin{eqnarray}
E_{O(2)} (\theta,\theta')= 
-\cos(\theta-\theta').
\label{EO(2)}
\end{eqnarray}
To compute the partition
function of this system, define a transfer ``matrix''
$$T(\theta,\theta')=e^{-\beta E(\theta,\theta')}.$$ 
Since the variables
of the system take continuous values, this isn't really a matrix, but
rather the kernel of an integral operator. It takes functions of
$\theta$ to functions of $\theta'$ by
\begin{eqnarray*}
g(\theta') = \int_0^{2\pi} \frac{d\theta'}{2\pi}\ T(\theta,\theta') f(\theta').
\end{eqnarray*}
To compute the partition function, we need eigenvalues $\lambda_l$
of $T$. Because the spins take values on a compact space (the circle here),
the eigenvalues are discrete and hence labeled by a discrete index $l$.
The corresponding eigenfunctions $f_l(\theta)$ obey
\begin{eqnarray}
\lambda_l f_l(\theta') = 
\int_0^{2\pi} \frac{d\theta'}{2\pi}\ T(\theta,\theta') f_l(\theta').
\end{eqnarray}
For the energy (\ref{EO(2)}), the $f_l(\theta)$ are obviously
$$f_l = e^{il\theta}.$$ 
The index $l$ must be an integer to preserve
the periodicity under $\theta\to \theta+2\pi$.
To see that these are eigenfunctions, note that
\begin{eqnarray*}
\lambda e^{il\theta}&=&
\int_0^{2\pi} \frac{d\theta'}{2\pi}\ e^{\beta\cos(\theta-\theta')} e^{il(\theta')}\\
&=&\int_0^{2\pi} \frac{d\theta'}{2\pi}\ e^{\beta\cos(\theta')} e^{il(\theta-\theta')}.\\
\end{eqnarray*}
The integral then can be evaluated for any $l$ in terms 
of a Bessel function:
\begin{eqnarray}
\lambda_l &=& 
\int_0^{2\pi} d\theta'\ e^{\beta\cos(\theta')} e^{-il\theta'}\cr\cr
&=& I_l(\beta).
\label{evo2}
\end{eqnarray}
The partition function for $L$ sites with periodic boundary conditions
is then
$$Z_{O(2)} = \sum_l  (I_l(\beta))^L.$$
When $L$ is large enough, the sum is dominated by largest eigenvalue, which 
here is the $l=0$ state. The internal energy density $U$ of the system
is then
$$U_{O(2)}= -\frac{1}{L}\frac{\partial}{\partial\beta} \ln Z =
-\frac{I'_0(\beta)}{I_0(\beta)}.$$ All other quantities such as
correlators can easily be found as well, since we have an explicit and
complete set of eigenvalues and their multiplicities.

\subsection{$O(N)$ magnets} 

The eigenvalues of the $O(2)$ problem are found by Fourier
transforming the transfer matrix. What we need to do for more
general cases can be summarized as Fourier analysis on manifolds more
general than the circle.  In other words, we want to
expand a function taking values on a manifold ${\cal M}$
into a series, e.g.
$$T({\cal M}) = \sum_l \lambda_l f_l({\cal M})$$ 
where the $f_l({\cal M})$ are complete set of orthonormal functions.
The eigenvalues of the transfer matrix are
the coefficients of the expansion in this basis.

The problem of Fourier analysis on all the manifolds of interest has
been solved already.  The key is to exploit the symmetry. A familiar
example is where the spins take values on the two-sphere $S^2$, where the
eigenfunctions are called spherical harmonics. In coset language, the
two-sphere can be described as the manifold $S^2 = O(3)/O(2)$: the
$O(3)$ group consists of rotations, while the $O(2)$ subgroup is the
set of rotations which leave a given point invariant. Thus different
points on $S^2$ take values in $O(3)/O(2)$. We parameterize the
two-sphere by the usual spherical coordinates: a unit three-vector
$\vec{s}\equiv
(\sin\theta\sin\phi,\sin\theta\cos\phi,\cos\theta)$.  To make
progress, it is crucial to consider an energy invariant under the
$O(3)$ rotation group, namely
$$E_{O(3)} = -\vec{s}\cdot {\vec{s'}} = -\cos\theta\cos\theta' -
\sin\theta\sin\theta'\cos(\phi-\phi').$$ 
We can expand the transfer
matrix energy into irreducible representations of the rotation group,
labeled by an angular momentum $l$ and an $L_z$ component $m$. 
The eigenfunctions of the transfer matrix are expressed in terms of
Legendre polynomials $P_{l}$, whose explicit definition will be given below.
One way of showing that the Legendre polynomials are
eigenfunctions of the transfer matrix is to show that they obey
an addition theorem, for example
\begin{eqnarray*}
&&P_{l}( \cos\theta\cos\theta' +
\sin\theta\sin\theta'\cos(\phi-\phi')) \\
&&\qquad\qquad = P_{l}(\cos\theta)
P_{l}(\cos\theta') + 2 \sum_{m=0}^l \frac{(l-m)!}{(l+m)!}
P_{lm}(\cos\theta) P_{lm}(\cos\theta') \cos(\phi-\phi'),
\end{eqnarray*} 
where the $P_{lm}$ are called associated Legendre polynomials,
with $P_l = P_{l0}$. One can then
expand the function $e^{\beta x}$ in terms of Legendre polynomials,
and then use the addition theorem to split the $\theta$ and $\theta'$
dependence. This leaves an integral for the eigenvalue.

To just obtain the eigenvalues of the transfer matrix, one does not
have to go to all the complications of generalized addition
theorems. Mathematicians have developed more in-depth ways of deriving
the eigenfunctions, and then the addition theorem comes as a
consequence of the computation. A geometric method is discussed in
\cite{Helgason}, while a much more explicit method is discussed in
\cite{Vilenkin}. We will require the methods of the latter in order to
treat the $Sp(N)$ magnet, where the spins do not take values on a
symmetric space. (A  symmetric space $G/H$ has $H$ a maximal subgroup of
$G$; the importance here is that when the spins take values
in a symmetric space, the transfer matrix depends on only one parameter.)

First we find the eigenvalues for an $O(N)$-invariant magnet, where
the spins $\vec{s}$ take values on the manifold $O(N)/O(N-1)$, which is the
$(N-1)$-sphere $S^{N-1}$.  We take the energy between nearest
neighbors to be
\begin{equation}
E_{O(N)} = - \vec{s}\cdot\vec{s'}.
\label{energyon}
\end{equation}
An eigenvalue $\lambda_l$ associated with eigenfunction $f_l(\vec{s})$
is given by the equation
\begin{equation}
\lambda^{O(N)}_l f_l(\vec{s}) = \int_{S^{N-1}} [D\vec{s'}]
e^{\beta \vec{s}\cdot\vec{s'}}
f_l(\vec{s'}).
\label{evon}
\end{equation}
where $[D\vec{s'}]$ is the usual measure on the $N-1$ sphere,
normalized so that $\int [D\vec{s'}] = 1$.  The unit $N$-vector
$\vec{s}$ depends on $N-1$ angles, but the energy only depends on the
angle between the two spins. This means that to compute the
eigenvalue, we need do only one integral. Put in terms of the $O(3)$
spherical harmonics, it means that the eigenvalues depend only the
value of $l$ and not $m$.  Explicitly, if one sets
$\vec{s}=(0,0,\dots,0,1)$, the integrand in (\ref{evon}) depends
on only a single angle $\theta$. We can then do the integral over
all the other angles in (\ref{evon}), and the measure
reduces to \cite{Vilenkin}
\begin{equation}
\int_{S^{N-1}} [D\vec{s'}] = \frac{\Gamma(N/2)}{\Gamma((N-1)/2)\Gamma(1/2)}
\int_0^\pi d\theta \sin^{N-2}\theta
\label{measure}
\end{equation}

The eigenfunctions for all the magnets we study here can be written in
terms of Jacobi polynomials $P^{(\mu,\nu)}_l(x)$. These are orthogonal
polynomials of order $x^l$, and a number of useful properties are
collected in the appendix.
The eigenfunctions for the $O(N)$ magnet with energy
(\ref{energyon}) are
given by $P^{(r,r)}(\cos\theta),$ where $r=(N-3)/2$. 
These are often called Gegenbauer or ultraspherical polynomials. They are indeed orthogonal
with respect to the measure (\ref{measure}). The eigenvalues are then given by
$$\lambda^{O(N)}_l  P^{(r,r)}(1)=
\frac{\Gamma(N/2)}{\Gamma((N-1)/2)\Gamma(1/2)}
\int_0^\pi d\theta \sin^{N-2}(\theta) e^{\beta\cos\theta}
P^{(r,r)}(\cos\theta) $$
Using the integral in the appendix for $\mu=\nu=(N-3)/2$ gives
$$
\lambda^{O(N)}_l = e^{-\beta} \frac{\Gamma(N/2)\Gamma(N/2-1/2 +l)}{\Gamma(1/2)
\Gamma(N-1+2l)} M(N/2-1/2+l,N-1+l,2\beta).$$
The function $M$ is called Kummer's function, and is a confluent
hypergeometric function. Its definition and the differential equation
it satisfies are given in the appendix.
Using the double-argument formula for gamma functions \cite{AS}
and the identity
(\ref{BesselM}) with $\nu=N/2-1+l$ gives our result
\begin{equation}
 \lambda^{O(N)}_l = \Gamma(N/2) \left(\frac{\beta}{2}\right)^{N/2-1+l}
I_{N/2 -1 +l}(\beta)
\end{equation}

Note that the eigenvalues reduce to the rotor result (\ref{evo2}) when
$N=2$.  Note also that $O(N)$ ferromagnets and antiferromagnets are
essentially the same, because redefining $\vec{s}\to -\vec{s}$ for
every other spin sends $\beta\to-\beta$. This transformation
$\theta\to \pi -\theta$ leaves the measure of the integral invariant,
and $\lambda^{O(N)}_l(-\beta)=(-1)^l \lambda^{O(N)}_l(\beta)$.

As a function of $\beta$, the Bessel function is analytic for all
$\beta$. Moreover $I_\gamma(\beta)>I_{\gamma+1}(\beta)$ for any
$\beta$ as long as $\gamma$ is positive. Thus the free energy does not
have any singularities as long as $\gamma$ remains finite. One might
hope something interesting happens when $N\to\infty$, but in Section
\ref{section:infinite-N} we will show that in this case there is still
no transition.

\subsection{$SU(N)$ magnets}

The computation for the $SU(N)$ case is very similar to that of the
$O(N)$ case, but we will see in the next section
how there is a completely new phase.

The $SU(N)$ magnet is defined in terms of a complex $N$-vector ${z}$
obeying $z^* \cdot z=1$. The energy for adjacent sites is 
\begin{equation}
E^{SU(N)} = - |z^* \cdot z'|^2. 
\label{esun}
\end{equation}
This energy is not
only invariant under global $SU(N)$ rotations, but is invariant under
local (gauge) transformations ${z}(x) \to e^{i\alpha(x)}{z}(x)$ at any
site $x$.
The vector ${z}$ takes values on a complex sphere $U(N)/U(N-1)$,
which as a manifold is identical to the real $2N-1$-sphere
$O(2N)/O(2N-1)$.  However, the $U(1)$ gauge symmetry can be used
to effectively reduce the number of degrees of freedom in the problem by
1. For example, one can set the last component of ${z}$ to be real
at every site, so ${z}$ effectively takes values on the manifold
$$\frac{U(N)}{U(N-1)\times U(1)}.$$ 
This manifold is a symmetric space, and is known as
the complex projective space $CP^{N-1}$. 

Like the $O(N)$ case, the energy between adjacent sites depends only
on a single angle $\theta$, where ${z}^*\cdot{z'} = \cos\theta$ (any
phase can always be gauged away).  Then the normalized measure can be
written as \cite{Vilenkin,Hikami}
\begin{equation}
\int_{CP^{N-1}} [D{z'}] = 2(N-1)
\int_0^{\pi/2} d\theta \sin^{2N-3}\theta \cos\theta 
\label{unmeasure}
\end{equation}
The eigenvectors of the $CP^{N-1}$ transfer matrix are Jacobi
polynomials as well, namely \cite{Vilenkin,Helgason}
$$P^{(N-2,0)}_l(\cos2\theta).$$
The $N=2$ case reduces to the ordinary real two-sphere, and indeed
the $P^{(0,0)}_l$ are ordinary spherical harmonics.
The $SU(N)$ symmetry means that the eigenvalues have degeneracies.
For example, we saw in section 2.1 that
for $N=2$ the eigenvalue depends only on $l$ and not the $S_z$ value $m$.
For $N=2$, the degeneracy $D_l$ of $\lambda_l$ is of course $2l+1$; the 
generalization to all $N$ is \cite{Vilenkin}
\begin{equation}
D_l = \frac{(2l+N-1)\Gamma^2(l+N-1)}{\Gamma(N)\Gamma(N-1)\Gamma^2(l+1)}
\label{degeneracy}
\end{equation}
The eigenvalues are given by
\begin{eqnarray} 
\lambda^{SU(N)}_l  P^{(N-2,0)}_l(1)&=& 2(N-1)
\int_0^{\pi/2} d\theta \sin^{2N-3}\theta \cos\theta\, e^{\beta\cos^2\theta}
P^{(N-2,0)}_l(\cos 2\theta) \cr\cr
&=& \frac{N-1}{2^{N-1}} \int_{-1}^1 dx (1-x)^{N-2} e^{\beta(1+x)/2} 
P^{(N-2,0)}_l(x)
\label{unevint}
\end{eqnarray}
where $x=\cos2\theta$.
Using the integral in the appendix for $\mu=N-2$, $\nu=0$ gives
\begin{eqnarray}
\lambda^{SU(N)}_l = \beta^{l} \frac{\Gamma(N)\Gamma(l+1)}{\Gamma(N+2l)}
M(l+1,N+2l,\beta)
\label{unev}
\end{eqnarray}
For finite $N$ and $\beta$, these are analytic functions.  Note that,
unlike for $O(N)$ magnets, $\beta>0$ and $\beta<0$ are not equivalent
here.  For positive $\beta$, we have a ferromagnet: the energy favors
aligned spins: a neighbor of $z = (0,0,\ldots,0,1)$ is in the same
state $z' = z$ (up to a phase).  However, $\beta<0$ does not
correspond to an antiferromagnet: the energy favors the neighbor $z'$
being in {\em any} state orthogonal to $z$, i.e., $z^* \cdot z' =
0$. There is a unique orthogonal state only for $N=2$.  Thus, with the
exception of $SU(2)$, the $SU(N)$ model (\ref{esun}) is ferromagnetic
when $\beta>0$, but is not an antiferromagnet when $\beta<0$.  For a
proper generalization of an antiferromaget, one needs to designate
among the $N-1$ vectors orthogonal to $z=(0,0,\ldots,0,1)$ one that
can be called ``antiparallel'' to it, for example,
$z'=(0,0,\ldots,1,0)$.  Doing so manifestly breaks the $SU(N)$
symmetry, as we discuss next.

\subsection{$Sp(N)$ magnets}

The $SU(2N)$ magnet can be deformed in an interesting way
by breaking the $SU(2N)$ symmetry to $Sp(N)$. 
The main novelty of this $Sp(N)$ magnet
is that the  phase transition can occur at an
antiferromagnetic value of $\beta$.
The energy between adjacent sites is
\begin{equation}
E^{Sp(N)} = - a\, |z^* \cdot z'|^2 - b\, |z J z'|^2.
\label{espn}
\end{equation}
where $z$ is a $2N$-dimensional complex vector, and $J$ is the
$2N\times 2N$ dimensional matrix $i\sigma_y\otimes I$, where I is the
$N$-dimensional identity matrix, and $\sigma_x$ the Pauli matrix.  
The second term breaks the $SU(2N)$ symmetry, but preserves an $Sp(N)$
subgroup.
The case $b=0$ reduces to the $SU(2N)$ ferromagnet discussed above.
The case $a=0$, studied by Read and Sachdev \cite{SR}, is a large-$N$
generalization of an antiferromagnet.  Indeed, the energy
\[
-|z J z'|^2 = 
-\left| \sum_{i=1}^N z_{2i-1} z'_{2i} + z_{2i} z'_{2i-1} \right|^2
\]
is minimized when vectors $z$ and $z'$ are locked in adjacent flavors
$2i-1$ and $2i$, e.g., $z=(0,0,\ldots,0,1)$ and $z'=(0,0,\ldots,1,0)$,
analogues of spin-down $(0,1)$ and spin-up $(1,0)$ in $SU(2)$. 

Like the $SU(N)$ magnet, the generalized model has a local $U(1)$
symmetry.  Just like the complex sphere $U(N)/U(N-1)$ is identical as
a manifold to the real sphere $O(2N)/O(2N-1)$, the ``quaternionic
sphere'' $Sp(N)/Sp(N-1)$ is identical as a manifold to the complex
sphere $U(2N)/U(2N-1)$ (which in turn is equivalent the real
$4N-1$-sphere $O(4N)/O(4N-1)$). When we fix a gauge, then $z$ takes
values on the manifold
\begin{equation}\frac{Sp(N)}{Sp(N-1)\times U(1)}.
\label{spnmanifold}
\end{equation}

It is easiest to 
study first the special point $a=b$, where the gauge symmetry is enhanced to
$SU(2)=Sp(1)$. The $SU(2)$ gauge symmetry (a subgroup of the $Sp(2N)$)
mixes the $2i-1$ and $2i$ components of $z$ and also mixes $z$ and
$z^*$. Precisely, if we arrange these components into the matrix
$$M_i =\pmatrix{z_{2i-1}&-z^*_{2i}\cr z_{2i}&z^*_{2i-1}} $$
then all the $M_i$ transform under the gauge symmetry as
$$M_i\to M_i U_i,$$ where $U_i$ is an element of $SU(2)$. The energy is
invariant under these transformations even though $U_i$ 
can be different at every point.
If we use this symmetry to fix a gauge (e.g.\
$z_{N-1}=\sin(\theta_{N-1})$, $z_{N}=\cos(\theta_{N-1})$),
then $z$ for $a=b$ takes values on the manifold
$$\frac{Sp(N)}{Sp(N-1)\times Sp(1)},$$
which we call $QP^{N-1}$.
Solving this case is almost the same as the $SU(N)$ case, because
this is a symmetric space.
The energy between adjacent sites depends on only one variable, and
the measure becomes \cite{Vilenkin,Hikami}
\begin{equation}
\int_{QP^{N-1}} [D{z'}] = 4(N-1)(2N-1)
\int_0^{\pi/2} d\theta \sin^{4N-5}\theta \cos^3\theta
\label{spnmeasure}
\end{equation}
The eigenfunctions
of the transfer matrix are the Jacobi polynomials
$P^{(2N-3,1)}(\cos2\theta)$ \cite{Vilenkin,Helgason}. 
Using the integral in the appendix yields
\begin{eqnarray}
\lambda^{QP}_l = \beta^{l} \frac{\Gamma(2N)\Gamma(l+2)}{\Gamma(2N+2l)}
M(l+2,2N+2l,\beta)
\label{spnev}
\end{eqnarray}
For $N=2$, this is equivalent to the $O(5)$ magnet.

The calculation for $a\ne b$ is more complicated. The reason is that
the energy between adjacent sites depends now on two angles:
the manifold (\ref{spnmanifold}) is not a symmetric space.
Luckily, the work of \cite{Vilenkin} allows us to solve
this case as well.
It is convenient to write coordinates
for the quaternionic sphere $Sp(N)/Sp(N-1)$ in terms 
of angles $\theta_i$ describing an ordinary $N-1$-dimensional sphere,
namely
\begin{eqnarray*}
q_1 &=& \sin\theta_{N-1}\dots\sin\theta_2 \sin\theta_1\\
q_2 &=& \sin\theta_{N-1}\dots\sin\theta_2 \cos\theta_1\\
q_3 &=& \sin\theta_{N-1}\dots\sin\theta_3 \cos\theta_2\\
\dots&&\\
q_{N-1} &=& \sin\theta_{N-1}\cos\theta_{N-2}\\
q_{N} &=& \cos\theta_{N-1}
\end{eqnarray*}
Then we have
\begin{eqnarray*}
z_{2i-1}&=&q_ie^{i\varphi_{2i-1}} \cos\omega_i \\
z_{2i}&=&q_i e^{i\varphi_{2i}} \sin\omega_i
\end{eqnarray*}
The angles take values in
$$0\le\theta_i\le \pi/2\qquad\quad 
0\le\omega_i\le \pi/2\qquad\quad 
0\le\varphi_i\le 2\pi.$$
Setting $z_i=(0,0,\dots,0,1)$ corresponds to all coordinates 
$\theta_i=\varphi_i=\omega_i=0$ except $\omega_N=\pi/2$. Then the energy
(\ref{espn}) is
$$ - (a\sin^2\omega'_N + b \cos^2\omega'_N) \cos^2\theta'_{N-1}$$
This indeed depends on two angles,
except in the $SU(2)$ gauge-invariant case $a=b$.
In terms of these two variables, the measure is
\begin{equation}
\int_{RS} [D{z}] = 8(N-1)(2N-1)
\int_0^{\pi/2} d\theta \sin^{4N-5}\theta \cos^3\theta
\int_0^{\pi/2} d\omega \sin\omega \cos\omega
\label{spnmeasure2}
\end{equation}
where we have dropped the now-unnecessary subscripts and primes.

The eigenvectors
of the corresponding transfer matrix are now labelled by two
indices, $0\le l'\le l < \infty$. They are 
written in terms of Jacobi polynomials as
\cite{Vilenkin}
$$ \cos^{l-l'}\theta\, P^{(0,0)}_{l-l'}(\cos 2\omega)
P^{(2N-3,l-l'+1)}(\cos 2\theta)$$
One can easily check that they are indeed orthogonal with
respect to the measure (\ref{spnmeasure2}). The eigenvalues
$\lambda_{ll'}$  of the
transfer matrix in the Read-Sachdev case
$a=0$, $b=1$ are therefore given by
\begin{eqnarray*}
&&\lambda^{RS}_{ll'} P^{(0,0)}_{l-l'}(1)
P^{(2N-3,l-l'+1)}(-1) =
8(N-1)(2N-1)
\int_0^{\pi/2} d\theta \sin^{4N-5}\theta \cos^3\theta\\
&&\qquad\qquad\qquad\qquad
\times \int_0^{\pi/2} d\omega \sin\omega \cos\omega
e^{\beta\cos^2\omega\cos^2\theta}
P^{(0,0)}_{l-l'}(\cos 2\omega)
P^{(2N-3,l-l'+1)}(\cos 2\theta)
\end{eqnarray*}
Using the integral in the appendix twice, we find that
$$
\lambda^{RS}_{ll'} = (-\beta)^{l-l'} \Gamma(2N) \sum_{j=0}^\infty
\beta^j A_j(l,l')
$$
where
$$A_j(l,l') = \frac{\Gamma(l-l'+1+j)
\Gamma(\frac{3}{2}l - \frac{3}{2}l' + 2 +j)
\Gamma(\frac{1}{2}l - \frac{1}{2}l' + 1 +j)}
{j!\,\Gamma(2l - 2l' + 2 +j)\Gamma(\frac{1}{2}l - \frac{3}{2}l' + 1 +j)
\Gamma(2N+\frac{1}{2}l +\frac{1}{2}l' + j)}
$$
This series be rewritten in terms of a confluent hypergeometric
function if desired. 
For the ground state $l=l'=0$, it simplifies to
$$\lambda^{RS}_{00} = M(1,2N,\beta).$$ The lowest eigenvalue is
identical to that for the $SU(2N)$ magnet, but the general eigenvalues
are not the same.

\section{Phase transitions as  $N\to\infty$}
\label{section:infinite-N}

In this section we study systems with the number of sites
$L\to\infty$. This means that the free energy follows from the largest
eigenvalue $\lambda_0$ of the transfer matrix. We saw in the last
section that in all cases, $\lambda_0$ is analytic function of the
inverse temperature $\beta$ for any finite $N$. Thus the only
possibility for a phase transition is if this function develops a
singularity as $N\to\infty$.

\subsection{No transition in the $O(N)$ magnet}

It is useful to study the $O(N)$ magnets first. We need
to use the asymptotic formula for Bessel functions, valid when
$\gamma$ is large \cite{AS}:
$$I_\gamma(\gamma y) \sim \frac{1}{\sqrt{2\pi\gamma}} 
\frac{e^{\gamma(\sqrt{1+y^2}-\xi^{-1})}}{(1+y^2)^{1/4}}$$
where
$$\xi^{-1} \equiv \ln(1+ \sqrt{1+y^2})-\ln(y)$$
This formula
implies that to have a non-trivial large $N$ limit, we need to take the
inverse temperature $\beta$ to $\infty$ as well, leaving the model in
terms of the new variable $y\equiv 2\beta/(N-2)$. Using this formula 
and some algebra gives the eigenvalue ratios to be
$$\frac{\lambda^{O(N)}_{l+1}(y)}{\lambda^{O(N)}_l(y)} = e^{-\xi^{-1}}$$
When we write quantities as a function of $y$, we mean that the expression is
valid up to terms of order $1/N$. The two-point function is
$$\langle\vec{s}_i\cdot\vec{s}_{i+R}\rangle \propto
\left(\frac{\lambda_1}{\lambda_0}\right)^R,$$ so $\xi$ is indeed the
correlation length.  At zero temperature ($y\to\infty$), $\xi$
diverges, just like the one-dimensional Ising model. This is the
usual zero-temperature behavior in one-dimensional classical models. The
correlation length does not diverge for any other value of
temperature, but note that at infinite temperature, $\xi\to 0$. This
is a state where the system is completely disordered: every site is
essentially independent of any other site, because all configurations
have the same weight in the partition sum. It is also useful to compute the 
internal energy. This is defined as
$$U=-\frac{\partial}{\partial y} \ln \lambda_0.$$
Using the asymptotic formula for the Bessel function gives 
$$\frac{U^{O(N)}(y)}{N} = \frac{1-\sqrt{1+y^2}}{y}$$ 
In the $N\to\infty$ limit, the 
internal energy is proportional to $N$, so it is the energy per component
which remains finite. 

\subsection{Transitions in the $SU(N)$ and $Sp(N)$ magnets}

The $SU(N)$ and $Sp(N)$ magnets each
have a phase transition when $N\to\infty$.
The lowest eigenvalue is given by for example
$$\lambda^{SU(N)}_0 = M(1,N,\beta).$$
As detailed in the last section, for the Read-Sachdev
case $\lambda^{RS}_{00} = \lambda^{SU(2N)}_0$.
To find the large $N$ behavior, it
is useful to examine the differential equation (\ref{diffeq}) for
$M(a,N,Ny)$ directly. Then one can see how
one can neglect various terms in the equation in various
regimes, and then easily solve the equation. 
This sort of analysis is called boundary-layer theory, and the
techniques are discussed at length in \cite{Bender}.
Rewriting (\ref{diffeq}) in terms of $y=\beta/N$, one has
\begin{equation}
\frac{y}{N} \frac{d^2 M}{dy^2} + (1-y) \frac{dM}{dy} - aM = 0.
\label{diffeqy}
\end{equation}
In the large $N$ limit, we can neglect the first term as long as $y<1$.
This shows that
\begin{equation}
\lim_{N\to\infty} M(a,N,Ny) = \frac{1}{(1-y)^a}\qquad\hbox{      for }y<1
.
\label{Mlim}
\end{equation}
This formula can be verified by
using Stirling's formula in the series expression (\ref{Kummer})
for $M$.
The lowest eigenvalue at large $N$ is therefore
\begin{eqnarray}
\lambda^{SU(N)}_0(y) \sim \frac{1}{1-y} \qquad\hbox{      for }y<1.
\label{evsunhigh}
\end{eqnarray}
Two important results are apparent
from this formula. First, there is a singularity at $y=1$. Second,
$U^{SU(N)}(y)$ for $y<1$ does not grow with $N$, and so the internal
energy per component vanishes. A vanishing internal
energy is characteristic of infinite temperature. The remarkable
characteristic of the $SU(N)$ magnet is that this behavior persists
all the way from infinite temperature $y=0$ to a finite temperature $y=1$.

In fact, the correlation length vanishes for all $y<1$. This follows from
the ratio of the first two eigenvalues:
\begin{eqnarray*}
\frac{\lambda^{SU(N)}_{1}(y)}{\lambda^{SU(N)}_0(y)} &=& (Ny)\frac{1}{N(N+1)}
\frac{M(2,N+2,Ny)}{M(1,N,Ny)}\\
&\sim& \frac{y}{N(1-y)}\hbox{    for }y<1.
\end{eqnarray*}
As $N\to\infty$, this vanishes if $y<1$.  Most scale-invariant
critical points have diverging correlation length. At the phase
transition at $y=1$, the correlation length {\it vanishes}.  We call a
phase with vanishing correlation length {\it seriously disordered}.
This distinguishes it from the $y>1$ phase, which is a conventional
disordered phase with a finite, non-zero correlation length.  In the
seriously-disordered phase, the eigenfunction is $P_0 = 1$. This
eigenfunction gives equal probabilities to all
configurations. Basically, what has happened is that the energy has
been swamped by the entropy. For large $N$, there are so many possible
spin configurations that if $y<1$, the energy term is too small to
make a difference. This is apparent in the integral
(\ref{unevint}). The measure favors configurations where the angle
$\theta$ between nearest-neighbor spins is near $\pi/2$. The reason is
as stated before: for a fixed ${z}$, there is only one value of ${z'}$
where ${z}^*\cdot {z'}=1$ but there are many with ${z}^*\cdot {z'}=0$.
The energy favors aligned spins, and at $y=1$ the energy term is
strong enough to cause a transition to a phase with finite correlation
length. The phase for $y>1$ is still a disordered phase, but a
conventional one, as occurs in the $O(N)$ magnets.  Note that this
phase transition occurs in the ferromagnetic phase ($\beta>0$). If
$\beta<0$, the energy and the entropy both favor disorder, so the
system is always seriously disordered.

We have asserted but not yet proven that for $y>1$, the internal
energy is non-zero and the correlation length is finite.  The
expression (\ref{Mlim}) for $M$ in the large-$N$ limit is not valid
for $y>1$, because the first term in the differential equation
(\ref{diffeqy}) can no longer be neglected when $(y-1)$ is of order
$1/\sqrt{N}$. To understand $y>1$, it is useful to derive a differential
equation for 
$$U=-\frac{\frac{d}{dy}M(1,N,Ny)}{M(1,N,y)}$$ directly. One has
\begin{eqnarray}
y\frac{d}{dy} U &=& -y\frac{M''}{M} + y \left(\frac{M'}{M}\right)^2\cr\cr
&=& U N(y-1)- N  + y U^2 
\label{diffeqU}
\end{eqnarray}
where we utilized (\ref{diffeqy}).
The internal energy for $y<1$ is a solution of this equation with
$U$ finite, found by neglecting the left-hand side and the last term
on the right-hand side. It is 
$$U^{SU(N)}(y) = \frac{1}{y-1}\qquad\qquad y<1$$
The solution of the differential equation (\ref{diffeqU})
for $N$ large and $y>1$ comes by assuming
$U/N$ is finite. Then one neglects the left-hand-side  and
the second term on the right-hand side, yielding
$$\frac{U^{SU(N)}(y)}{N} = \frac{1-y}{y}\qquad\qquad y>1$$
One can find the $1/N$ corrections to these expressions systematically.
For example, one can show that
the lowest eigenvalue $\lambda_0$ for $y>1$ is to next order in $N$:
\begin{equation}
\lambda_0^{SU(N)}(y) = A_N e^{N(y-1)} y^{1-N} \qquad\qquad y>1
\label{evsunlow}
\end{equation}
where $A_N =\sqrt{2\pi N}$.

The ratio of the first two eigenvalues for $y>1$ follows from a
similar computation, yielding
\begin{eqnarray*}
\frac{\lambda^{SU(N)}_{1}(y)}{\lambda^{SU(N)}_0(y)} &=& \frac{y}{N+1}
\frac{M(2,N+2,Ny)}{M(1,N,Ny)}\\
&\sim& \frac{y-1}{y}\qquad\qquad y>1
\end{eqnarray*}
Thus we see that the correlation length indeed goes to zero as $y\to
1$ from above.
This is a second-order phase transition: the energy is continuous but
its derivative is discontinuous. 
For the Read-Sachdev $Sp(N)$ magnets, the formula is similar:
\begin{eqnarray*}
\frac{\lambda^{RS}_{1}(y)}{\lambda^{RS}_0(y)}
&\sim& \sqrt{\frac{y-1}{y}}\qquad\qquad y>1
\end{eqnarray*}

\subsection{Near the phase transition}

The expression for the eigenvalues in terms of the Kummer function $M$
is valid for any value of $y$ and $N$, but the expressions derived for
the large-$N$ limit break down when $|y-1|$ is of order $(\sqrt{N})$.
All the terms in the differential equations
(\ref{diffeqy},\ref{diffeqU}) need to be included in this region. 

To understand the  transition region, we study the physics in terms
of the variable
$$ x\equiv \sqrt{N}(y-1).$$
The differential equation for the lowest eigenvalue
(\ref{diffeqy}) becomes (in the $SU(N)$ case)
\begin{equation}
\left(1+\frac{x}{\sqrt{N}}\right) \frac{d^2 \lambda_0}{dx^2} - x \frac{d\lambda_0}{dx} - \lambda_0 = 0.
\label{diffeqx}
\end{equation}
For $N$ large, we can neglect the term with the $1/\sqrt{N}$ in it. Thus
the subsequent analysis will be valid up to terms of order $1/\sqrt{N}$,
as opposed to equations in the last section, which have corrections of
order $1/N$. Solving this differential equation by plugging in the
series
$$\lambda_0= \sum_{j=0}^\infty m_j x^j,$$
requires that 
$$m_{j+2}=\frac{1}{j+2}m_j$$
Plugging this in and summing the series,
the solution near $y-1$ for large $N$ is therefore
\begin{equation}
\lambda_0=m_0 e^{x^2/2} + m_1 x M(1,3/2,x^2/2)
\label{lam0}
\end{equation}
where the $M$ is indeed our friend the Kummer function. 
Thus we have shown that the eigenvalue $\lambda_0= M(1,N,Ny)$
near $y-1$ at large $N$ is related to a
Kummer function with different arguments.
To fix the values of $m_0$ and $m_1$, we need to match this onto
the value of $\lambda_0$ in the high-temperature phase (\ref{evsunhigh})
and in the low-temperature phase (\ref{evsunlow}). From \cite{AS}, we have
the asymptotic formula
$$ M(1,3/2,z) \approx \frac{1}{2}\sqrt{\frac{\pi}{z}}
e^z  \qquad \qquad
z\hbox{ large}$$
Matching while neglecting terms of order $1/\sqrt{N}$ yields
$m_0 = m_1\sqrt{\pi/2}$. Matching this with the numerical result
for $A_N$ given in the last section gives
$$m_0 = \sqrt{\frac{\pi N}{2}}\qquad \qquad m_1= \sqrt{N}$$
The two terms in (\ref{lam0}) cancel when $x$ is large 
and negative, but add when $x$ is large and positive.

With a little work, one can now derive the specific heat
$$C= \frac{dU}{d(1/y)}$$
everywhere, including at $y=1$.
We find $C=1$ for $y<1$, $C=0$ for $y<1$, and $C=1-(2/\pi)$ for $y=1$.
The traditional critical exponents can be defined when $N=\infty$. For the
specific heat, we have $\alpha=0$, while the correlation
length goes to zero logarithmically, so $\nu=0$ as well.

\section{First-order transition for three sites}

In section 2 we found all the eigenvalues and their degeneracies for
the $SU(N)$ magnet. This means that we can also study phase transitions
where the number of sites $L$ is finite as well as the infinite-$L$ case
studied in section 3. The phase transition persists
all the way down to two sites. A remarkable result
of the mean-field-theory analysis of \cite{TS}
is that for 
three sites (and only for three sites), the transition becomes first-order.
We verify this result in this section.

The partition function for the $SU(N)$ magnet
with periodic boundary conditions for any number of sites $L$ is
given by
$$Z= \sum_{l=0}^\infty D_l (\lambda_l)^L$$ where the eigenvalues
$\lambda_l$ and their degeneracies $D_l$ are given by (\ref{unev}) and
(\ref{degeneracy}) respectively.  In this section we study only the
$SU(N)$ magnet, so we will omit the $SU(N)$ superscripts. If one takes
$L\to\infty$ before taking $N\to\infty$, the partition function is
dominated by the largest eigenvalue $\lambda_0$ no matter what the
degeneracy $D_l$ is. However, for finite $L$, the fact that $D_l$
grows quite quickly with $l$ and $N$ means that the largest eigenvalue
does not necessarily dominate the partition function, but instead
$$Z\approx D_{l_0} \lambda_{l_0}^L$$
for some $l_0$ which may not be zero.

All the eigenvalues with finite $l$ (so that $l/N\to 0$ as
$N\to\infty$) have the same kind of singularity as $y\to 1$. Thus if
the effect of including the degeneracy merely shifts $l_0$ to some
finite value, the second-order transition remains.  To find a
first-order transition, we need to study the
eigenvalues when $N\to\infty$ with $l/N$ remaining finite. 
It is convenient to fix the
inverse temperature $y$ and study the problem as a function of $r$,
defined as
$$r=2\frac{l}{N}.$$
At some value
$r=r_0$, $D(r)\lambda(r,y)^L$ is a maximum; at this value
the free energy is minimized (the path integral has a saddle point).
The behavior of the eigenvalues at non-zero $r$ can be found by
doing a saddle-point approximation to the integral representation
of the Kummer function, or by deriving a differential equation for $M'/M$
in the manner of (\ref{diffeqU}). To leading order in $N$,
\begin{eqnarray}
\frac{2}{N} \ln(\lambda(r,y))&=&
A+y-1+\ln(2)+r\ln(y)+(r+1)\ln(r)\cr\cr
&&\qquad -\ln(y-1+A)-(r+1)\ln\left((r+1)^2-y +(r+1)A\right)
\label{lamr}
\end{eqnarray}
where
$$A=\sqrt{(r+1)^2 + (y-1)^2 -1}.$$
At large $N$ and $l$, the degeneracy behaves
as
\begin{eqnarray}
\frac{1}{N} \ln(D(r)) =(r+2)\ln(r+2)-r\ln(r) - 2\ln(2)
\end{eqnarray}
The degeneracy is a group-theoretical factor and is independent of the
inverse temperature $y$.

\bigskip
\begin{figure}[here]
\centerline{\includegraphics[scale=1.0]{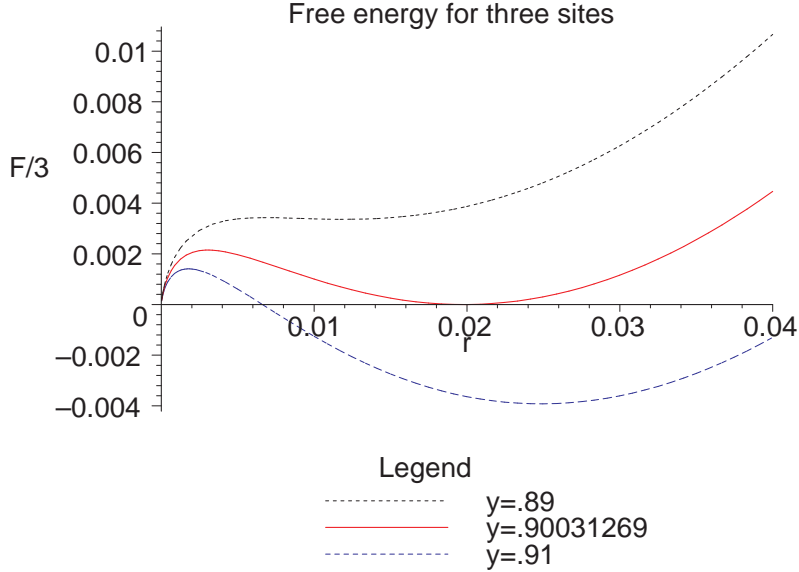}}
\bigskip\bigskip
\caption{Plots of the free energy per site at and near the first-order
transition for three sites}
\label{fig:free-energy}
\end{figure}
A first-order transition occurs if
$r_0$ jumps discontinuously as $y$ is varied. What can happen is
that $D(r)\lambda(r,y)^L$ can develop another maximum as a function of
$r$. At some value $y=y_0$, one can have two peaks at the same place:
\begin{eqnarray*}
D(r_1)\lambda(r_1,y_0)^L &=& D(r_2)\lambda(r_2,y_0)^L\\
\frac{d}{dr}\left[D(r)\lambda(r,y_0)^L\right]_{r=r_1} 
&=& \frac{d}{dr} \left[D(r)\lambda(r,y_0)^L\right]_{r=r_2}=0.
\end{eqnarray*}
If this happens, the free energy has two
minima as a function of $r$. 
This is the mark of a first-order transition: the internal energy will
not be continuous in $y$ because the value of $r_0$ jumps from $r_1$
to $r_2$ as $y$ is varied.  We find that for $L=3$, this indeed
happens.  In Fig.\ \ref{fig:free-energy}, we plot the free energy
per site $F/3 = -\ln{[D(r)\lambda(r,y)^3]}/(3y)$ for values of $y$
just below, above, and at the transitional value
$y_0=.900312694913\dots$. At the phase transition, the value of
$r_0$ jumps from $0$ to $.0197950036795\dots$. This is small, but the transition is definitely
first order, as shown in \cite{TS}.
The transition occurs for $y<1$, so the internal
energy jumps from $0$ to a finite value. No transition occurs at $y=1$,
because the saddle point is already away from $r=0$.

For any other value of $L$, the transition is second-order as before;
no second minimum seems to occur. Due to the unwieldiness of the
expression (\ref{lamr}), we have not been able to prove this in
general, but it is easy to see by looking at the curves numerically.
Since the first-order transition is so weak for three sites, it would
indeed be surprising if the effect of the degeneracies were to
overcome the behavior of the eigenvalues for larger $L$.  To see this
in more detail, let us examine some limits. At large $r$, the
eigenvalue goes to zero as $(yr)^{-Nr}e^{Nr}$, while the degeneracy
only grows as $r^{2N}$. Thus $r_0$ must be finite.  The behavior for
small $r$ crucially depends on whether $y$ is greater or less than
$1$. For $y<1$, $D(0)\lambda(y,0)^L =1$, and the slope at $r=0$ is
negative. For $y>1$, $D(0)\lambda(y,0)^L$ depends on $y$ and the slope
at $r=0$ is positive.  The simplest behavior for $y<1$ consistent with
these limits is to have $D(r)\lambda(y,r)^L$ fall off monotonically
from $1$ to zero. The simplest behavior for $y>1$ is for
$D(r)\lambda(y,r)^L$ to rise up to a single maximum at some value
$r=r_0$, and then fall off to zero for large $r$. 
By studying plots of the free energy, it
seems that this simple scenario is realized for all $L$ larger than
$3$.
For $L>3$ and $y<1$, $r_0=0$, so the partition function is
dominated by an eigenvalue with finite $l$. As $y$ is increased past
$1$, $r_0$ becomes non-zero.  There is only a single peak, so $r_0$
varies continuously with $y$, so the minimum value of the free energy
is also varies continuously.
This means that for $L>3$
the only phase transition is a second-order one at $y=1$.

\bigskip\bigskip We are very grateful to Shivaji Sondhi for many
interesting conversations and for collaboration on \cite{TS}. We are
also grateful to R.~Moessner and N.~Read for helpful conversations.
The work of P.F.\ is supported by NSF Grant DMR-0104799, a DOE OJI
Award, and a Sloan Foundation Fellowship.  O.T.\ is supported by 
NSF Grant DMR-9978074.  

\appendix
\section{Jacobi polynomials and confluent hypergeometric functions}

The Jacobi polynomials $P^{(\mu,\nu)}(x)$
can be expressed in terms of a Rodrigues formula
$$(1-x)^\mu (1+x)^\nu P^{(\mu,\nu)}_l(x) = \frac{(-1)^l}{2^l l!}
\left(\frac{d}{dx}\right)^l\left[(1-x)^{l+\mu} (1+x)^{l+\nu}\right]$$
They are defined to be orthogonal with respect to the measure $dx
(1-x)^\mu (1+x)^\nu$:
$$\int_{-1}^1 dx\, (1-x)^\mu (1+x)^\nu  P^{(\mu,\nu)}_l(x)
 P^{(\mu,\nu)}_k(x)=0 \qquad\qquad l\ne k$$
Many results on Jacobi polynomials can be found in \cite{Szego}.
One integral we need is \cite{Bateman}
$$
\int_{-1}^1 dx\, (1-x)^\mu (1+x)^\sigma  P^{(\mu,\nu)}_l(x) =
\frac{2^{\mu+\sigma +1} \Gamma(\sigma + 1) \Gamma(\mu+l+1)
\Gamma(\sigma-\nu +1)}{l!\ \Gamma(\sigma-\nu-l+1)\Gamma(\mu+\sigma+l+2)}
$$
Be aware that there are (different) typos in this formula in both
\cite{Bateman} and in \cite{GR}.
Another useful result is
$$P^{(\mu,\nu)}(1) = 
\frac{\Gamma(l+\mu+1)}{\Gamma(l+1)\Gamma(\mu+1)}$$
Using these two relations we evaluate the integral
\begin{eqnarray*}
&&\int_{-1}^1 dx\, (1-x)^\mu (1+x)^\nu e^{\beta (1+x)} P^{(\mu,\nu)}_l(x)\\ 
&&\qquad\qquad=
\sum_{j=0}^\infty \frac{\beta^j}{j!}
\int_{-1}^1 dx\, (1-x)^\mu (1+x)^{\nu +j}  P^{(\mu,\nu)}_l(x)
 \\
&&\qquad\qquad=
\sum_{j=0}^\infty \beta^j 
2^{\mu+\nu+j +1} \frac{\Gamma(\nu+j + 1) \Gamma(\mu+l+1)}
{l!\,\Gamma(j-l+1)\Gamma(\mu+\nu+j+l+2) }\\
&&\qquad\qquad=
P^{(\mu,\nu)}_l(1)\,\Gamma(\mu+1) \,\beta^l\, \sum_{k=0}^\infty
(2\beta)^k 2^{\mu+\nu+l +1} \frac{\Gamma(\nu+k + l+1)}
{k!\,\Gamma(\mu+\nu+k+2l+2)} \\
&&\qquad\qquad=
P^{(\mu,\nu)}_l(1)\,\beta^l\,2^{\mu+\nu+l+1}\frac{\Gamma(\mu+1)\Gamma(\nu+l+1)}{\Gamma(\mu+\nu+2l+2)} M(\nu+l+1,\mu+\nu+2l+2,2\beta) 
\end{eqnarray*}
where $k=j-l$ and we use the fact that $1/\Gamma(j-l+1) = 0$ for $j-l$ a negative integer.
The function $M(a,b,z)$ is known as Kummer's series:
\begin{equation}
M(a,b,z) = \frac{\Gamma(b)}{\Gamma(a)}\sum_{k=0}^\infty \frac{z^k}{k!}
\frac{\Gamma(a+k)}{\Gamma(b+k)}.
\label{Kummer}
\end{equation}
The Kummer function is a confluent hypergeometric function
because it can be obtained
by taking a limit of the hypergeometric function ${}_2F_1$ where
two singularities coincide. It satisfies the differential
equation
\begin{equation}
z \frac{d^2 M}{dz^2} + (b-z) \frac{dM}{dz} - aM = 0
\label{diffeq}
\end{equation}
$M$ is an analytic function of $a,b$ and $z$; the only way to get
singularities is to take some or all of these parameters to infinity.
A fact useful for $O(N)$ magnets is that
the Bessel function $I_\gamma(z)$ can be written as
\begin{equation}
I_\gamma(z) = \frac{z^\nu}{2^\nu\,\Gamma(\gamma+1)} e^{-z}
M(\gamma+1/2,2\gamma+1,2z)
\label{BesselM}
\end{equation}

\end{document}